\documentclass[aip,jcp,reprint,onecolumn,amsmath,amssymb,amsfonts,preprintnumbers]{revtex4-1}
\usepackage{bm,graphicx,xcolor,microtype,multirow}

\newcommand{\Ec}{\epsilon_{\rm c}}

\newcommand{\kF}{k_{\text{F}}}

\newcommand{\alert}[1]{\textcolor{black}{#1}}

\begin{document}

\title{High-density correlation energy expansion of the one-dimensional uniform electron gas}

\author{Pierre-Fran\c{c}ois Loos}
\email{loos@rsc.anu.edu.au}
\affiliation{Research School of Chemistry, 
Australian National University, Canberra, ACT 0200, Australia}

\begin{abstract}
We show that the expression of the high-density (i.e small-$r_s$) correlation energy per electron for the one-dimensional uniform electron gas can be obtained by conventional perturbation theory and is of the form $\Ec(r_s) = -\pi^2/360 + 0.00845\,r_s + \ldots$, where $r_s$ is the average radius of an electron. Combining these new results with the low-density correlation energy expansion, we propose a local-density approximation correlation functional, which deviates by a maximum of 0.1 millihartree compared to the benchmark DMC calculations.
\end{abstract}

\maketitle

\section{Introduction}

Recently much attention has been devoted to  one-dimensional (1D) systems. For example, Wagner {\em et al.} \cite{Wagner12} have shown that 1D chemical systems, such as light atoms (H, He, Li, Be, \ldots), ions (H$^{-}$, Li$^{+}$, Be$^{+}$, \ldots) or diatomics (e.g. H$_2$),  can be used as ``theoretical laboratory'' to study strong correlation in ``real'' three-dimensional chemical systems within density-functional theory (DFT). \cite{ParrBook} 

One-dimensional systems can be also experimentally realized in carbon nanotubes, \cite{SaitoBook, Egger98, Bockrath99, Ishii03, Shiraishi03} organic conductors, \cite{Schwartz98, Vescoli00, Lorenz02, Dressel05, Ito05} transition metal oxides, \cite{Hu02} edge states in quantum Hall liquids, \cite{Milliken96, Mandal01, Chang03} semiconductor heterostructures, \cite{Goni91, Auslaender00, ZaitsevZotov00, Liu05, Steinberg06} confined atomic gases, \cite{Monien98, Recati03, Moritz05} and atomic or semiconducting nanowires. \cite{Schafer08, Huang01} The uniform electron gas (UEG) paradigm which is  the main ``ingredient'' of most of the correlation functionals and the cornerstone of the most popular DFT approximation---the local-density approximation (LDA)--- is particularly well-adapted to the theoretical study of subtile effects involved by electron correlation in such systems. However, while the high-density (small-$r_s$) reduced (i.e. per electron) correlation energy expansions
\begin{equation}
\label{Ecjellium}
\begin{split}
	\Ec(r_s) 
	& = \sum_{j=0}^\infty \left(\lambda_j \ln r_s + \epsilon_j\right)r_s^j
	\\
	& = \lambda_0 \ln r_s + \epsilon_0 + \lambda_1 r_s \ln r_s + \epsilon_1 r_s + \ldots
\end{split}
\end{equation}
(where $r_s$ is the Seitz radius) of the two-dimensional (2D) and three-dimensional (3D) UEGs are quite well-known, \cite{Zia73, Isihara77, Rajagopal77, Glasser77, Isihara80, Glasser84, Seidl04, Giuliani07, 2DEG, Macke50, Bohm53, Pines53, GellMann57, DuBois59, Carr64, Misawa65, Onsager66, Wang91, Hoffman92, Endo99, Ziesche05, Sun10, 3DEG, Handler12, Glomium11} much less has been discovered about the 1D UEG. This lack of information is mainly due to the divergence of the Coulomb operator $1/x$ in 1D for small interelectronic distance $x$, \cite{Drummond07, Astrakharchik11, Lee11a, QR12, nQR12} which makes conventional perturbation theory difficult to apply due to the absence of a Fourier transform for the Coulomb operator. In this Regular Article, we propose to fill this gap by reporting the values of the first few high-density coefficients (see Table \ref{tab:coeffs}). \alert{We note that, although the bare Coulomb operator is not the natural operator in 1D (i.e. the solution of the 1D Poisson's equation does not give a Coulombic potential), in the following study we are interested in real electrons that are confined so that they can move in only one dimension of a 3D space.  For this reason, it is appropriate to use the 1/x Coulomb potential.}

The present system is constructed by allowing the number $n$ of electrons in a 1D box of length $L$ with periodic boundary conditions to approach infinity with the density
\begin{equation}
	\rho = \frac{n}{L}=\frac{1}{2r_s}
\end{equation}
 held constant. \cite{Vignale, ParrBook} Because the paramagnetic and ferromagnetic states are degenerate for strict 1D systems, we will consider only the latter (i.e. a spin-polarized electron gas). \cite{Lee11a, QR12, nQR12} 

To avoid the divergence of the Coulomb operator, we will consider in our derivation a ``soften'' version of the Coulomb operator $1/\sqrt{x^2+R^2}$, where $R$ is a parameter which removes the singularity at $x=0$. \cite{Schulz93, Fogler05a}  Then, we will carefully take the limit $R \to 0$. We will show that, unlike the 2D and 3D version of the UEG, second- and third-order perturbation theories are convergent, i.e there is no need to use resummation techniques. \cite{GellMann57} Combining these new results with the low-density energy expansion and the available diffusion Monte Carlo (DMC) data, we propose a new LDA functional for the reduced correlation energy of the 1D UEG. Atomic units are used throughout. 

\section{High-density expansion}

\subsection{Second-order perturbation theory}

In 1D, the spinorbitals of the free electron gas are
\begin{equation}
	\psi_k(x) = \frac{e^{i k x}}{\sqrt{L}},
\end{equation}
with the energy $\kappa_k = k^2/2$, and where the periodic boundary conditions imply $k=2\pi\,m/L$ ($m \in \mathbb{Z}$). The coefficient $\epsilon_0$ is given by second-order perturbation theory \cite{SzaboBook}
\begin{equation}
	\epsilon_0(R) = \frac{1}{4n} \sum_{ab}^{\text{occ}} \sum_{rs}^{\text{virt}} \frac{\left| \langle ab\Vert rs \rangle \right|^2}{\kappa_a + \kappa_b - \kappa_r - \kappa_s},
\end{equation}
where $\langle ab\Vert rs \rangle = \langle ab \vert rs \rangle - \langle ab \vert sr \rangle$ and
\begin{equation}
	\langle ab \vert rs \rangle =\int_{-L/2}^{L/2} \int_{-L/2}^{L/2} \frac{\psi_a^*(x_1)  \psi_b^*(x_2) \psi_r(x_1)  \psi_s(x_2)}{\sqrt{(x_1-x_2)^2+R^2}} dx_1 dx_2.
\end{equation}
The constant coefficient $\epsilon_0$ is usually decomposed into a direct (``ring-diagram'') term $\epsilon_0^\text{a}$ and an exchange term $\epsilon_0^\text{b}$, which read explicitly as
\begin{align}
	\label{eps0-a}
	\epsilon_0^\text{a}(R)  & = \frac{1}{2n} \sum_{ab}^{\text{occ}} \sum_{rs}^{\text{virt}} \frac{\langle ab | rs \rangle\langle rs | ab \rangle}{\kappa_a + \kappa_b - \kappa_r - \kappa_s},
	\\
	\label{eps0-b}
	\epsilon_0^\text{b}(R)  & = - \frac{1}{2n}  \sum_{ab}^{\text{occ}} \sum_{rs}^{\text{virt}}\frac{\langle ab | rs \rangle\langle rs | ba \rangle}{\kappa_a + \kappa_b - \kappa_r - \kappa_s}.
\end{align}
Using the Fourier transform of the soft Coulomb potential 
\begin{equation}
	\frac{1}{\sqrt{x^2+R^2}} = \frac{1}{\pi} \int_{-\infty}^{\infty} K_0(|k|R)e^{i k x} dk,
\end{equation}
where $K_0$ is the zeroth-order modified Bessel function of the second kind, \cite{NISTbook} the well-known relation
\begin{equation}
	\delta(x) = \frac{1}{2\pi} \int_{-\infty}^{\infty} e^{i k x} dk,
\end{equation}
and transforming the sums in \eqref{eps0-a} and \eqref{eps0-b}  into integrals 
\begin{equation}
	\sum_k \rightarrow  \frac{L}{2\pi}  \int dk,
\end{equation}
with $p_1 = a/\kF$, $p_2 = b/\kF$, $q = k/\kF$ and $R \leftarrow \kF R$, where $\kF=\pi \rho$ is the Fermi wave vector, \cite{RaimesBook} we eventually find \footnote{Energies in Raimes' book \cite{RaimesBook} are reported in Rydberg, while here we use atomic units.}
\begin{equation}
	\label{eps-a-final}
	\epsilon_0^\text{a}(R) = - \frac{1}{4\pi^2}  \int_{-\infty}^{\infty} dq \int_{\substack{|p_1|<1\\|p_1+q|>1}} dp_1 
	\int_{\substack{|p_2|<1\\|p_2+q|>1}} dp_2 \frac{K_0(|q|R)^2}{q(p_1+p_2+q)},
\end{equation}
and
\begin{equation}
	\label{eps-b-final}
	\epsilon_0^\text{b}(R) = \frac{1}{4\pi^2}  \int_{-\infty}^{\infty} dq \int_{\substack{|p_1|<1\\|p_1+q|>1}} dp_1  
	\int_{\substack{|p_2|<1\\|p_2+q|>1}} dp_2 \frac{K_0(|q|R)K_0(|p_1+p_2+q|R)}{q(p_1+p_2+q)}.
\end{equation}

\begin{table*}
\caption{\label{tab:coeffs} 
High-density coefficients for the paramagnetic state of the 1D, 2D and 3D UEGs. The paramagnetic and ferromagnetic states are degenerate in 1D. $\beta$ and $\zeta$ are the Dirichlet beta and Riemann zeta functions, respectively. \cite{NISTbook}}
\begin{ruledtabular}
\begin{tabular}{cccccc}
	Coefficient				&	Term					&	1D										&	2D												&	3D																\\
	\hline
	$\lambda_0$				&	$\ln r_s$				&	$0$										&	$\displaystyle 0$									&	$\displaystyle (1-\ln2)/\pi^2$											\\
	$\epsilon_0$				&	$r_s^0$				&	$\displaystyle -\pi^2/360$						&	$\displaystyle \ln 2 - 1+\beta(2)-8\beta(4)/\pi^2$			&	$\displaystyle -0.071099+(\ln 2)/6-3\zeta(3)/(4\pi^2)$						\\
	$\lambda_1$				&	$r_s\ln r_s$			&	$0$										&	$\displaystyle - \sqrt{2} \left(10/(3\pi)-1\right)$				&	$\displaystyle +0.009229$											\\
	$\epsilon_1$				&	$r_s$				&	$+0.00845$								&	unknown											&	$-0.020$															\\
\end{tabular}
\end{ruledtabular}
\end{table*}

\begin{figure}
\includegraphics[width=0.4\textwidth]{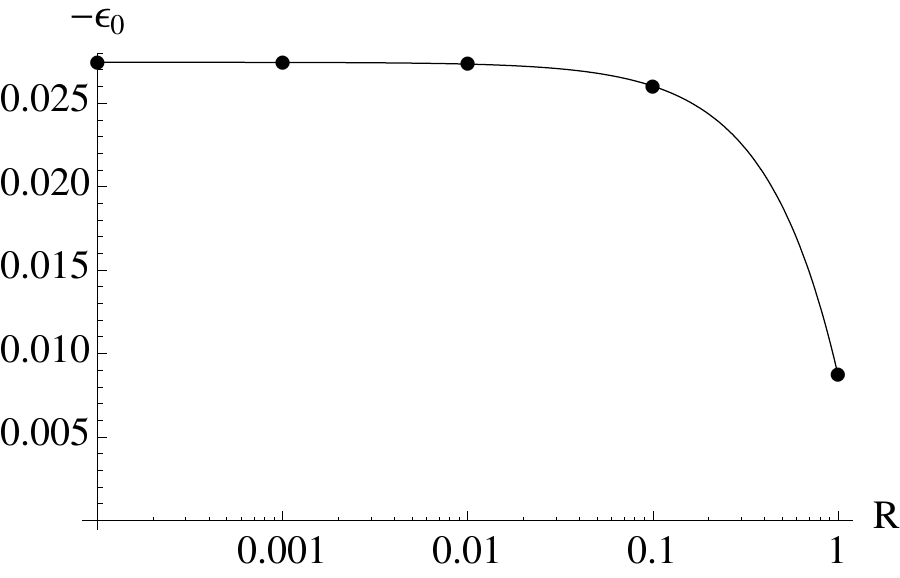}
\includegraphics[width=0.4\textwidth]{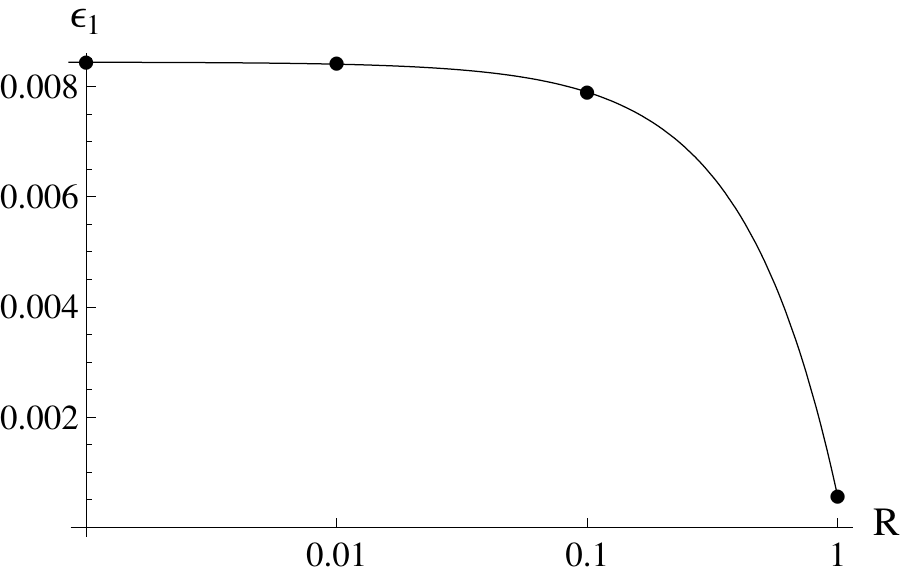}
\caption{
\label{fig:conv}
Convergence of $\epsilon_0(R)=\epsilon_0^\text{a}(R)+\epsilon_0^\text{b}(R)$ (Eqs.~\eqref{eps-a-final} and \eqref{eps-b-final}) and $\epsilon_1(R)=\pi^{-4} \sum_{k=1}^{8} \Xi_i(R)$ (explicit expressions given in Table \ref{tab:epsilon1}) with respect to $R$.}
\end{figure}

For $R>0$, Eqs.~\eqref{eps-a-final} and \eqref{eps-b-final} can be evaluated numerically. As shown in Fig.~\ref{fig:conv},  $\epsilon_0(R)$ decreases monotonically to reach a constant at $R=0$. However, for $R=0$, both integral diverge at opposite rates. Thus, to find the limiting value, it is better not to split $\epsilon_0(R)$ into two contributions but to consider them together. For small $R$, we have
\begin{equation}
	\epsilon_{0}(R) = - \frac{1}{2\pi^2} \int_{0}^{\infty} dq \int_{\substack{|p_1|<1\\|p_1+q|>1}} dp_1 
	 \int_{\substack{|p_2|<1\\|p_2+q|>1}} dp_2 
	\frac{[\ln q  - \ln(p_1+p_2+q)][\ln (q R/2)+ \gamma]}{q(p_1+p_2+q)} + O(R^2).
\end{equation}
The integrations over $p_1$ and $p_2$ can be performed at this stage and it yields
\begin{equation}
	\epsilon_{0}(R) = - \frac{1}{4\pi^2} \int_{0}^{\infty}\frac{ \Lambda(q)}{q} [\ln (q R/2)+ \gamma]dq,
\end{equation}
where $\gamma$ is the Euler-Mascheroni constant \cite{NISTbook} and
\begin{equation}
\begin{split}
	\Lambda(q) & = (2 + q) [\ln(2 +q)]^2 + (2 - q) [\ln(2 - q)]^2 
	\\
	& - 2(1+ \ln q) (2 +q) \ln(2 + q) 
	\\
	& -  2 (1 +  \ln q) (2 - q)  \ln(2 - q)
	\\
	& - 4 (\ln 2)^2 + 8 \ln 2 (1 + \ln q)
\end{split}
\end{equation}
for $0 \leq q \leq 2$, and 
\begin{equation}
\begin{split}
	\Lambda(q) & = (2 + q) [\ln(2 + q)]^2 + (2 - q) [\ln(q-2)]^2 
	\\
	& - 2(1+ \ln q) (2 + q) \ln(2 + q) 
	\\
	& -  2 (1 +  \ln q) (2 - q)  \ln(q-2) 
	\\
	& + 2 (2+\ln q) q \ln q 
\end{split}
\end{equation}
otherwise. Performing the last integration over the two distinct regions ($0 \leq q \leq 2$ and $q>2$) gives two contributions that diverges as $\ln R$ for small $R$ with opposite sign.
Thus, the divergences cancel and we find
\begin{equation}
	\epsilon_0 = \lim_{R \to 0} \epsilon_0(R) = -\frac{\pi^2}{360},
\end{equation}
which nicely reproduces the result obtained with a ring geometry. \cite{nQR12} 

Because $\Ec(r_s) = \epsilon_0 + O(r_s)$ (see below), $\epsilon_0$ provides the \textit{exact} value of the correlation energy at $r_s=0$, and it is roughly $-27.4$ millihartree per electron. It is worth noting that, in most of the studies on 1D systems, a soft Coulomb operator is considered. For example, in Refs.~\onlinecite{Wagner12} and \onlinecite{Stoudenmire12}, the authors used $R=1$, yielding a correlation energy (-8.7 millihartree) more than three times smaller than the value obtained using the genuine Coulomb operator (i.e $R=0$). Moreover, using a quasi-1D model with a transverse harmonic potential, Casula et al. conclude that, in the high-density limit, the correlation energy vanishes quadratically with $r_s$. \cite{Casula06} This strikingly different prediction stresses the importance of employing a realistic Coulomb operator

\subsection{Third-order perturbation theory}

Using the same approach, third-order perturbation theory gives \cite{Carr64}
\begin{equation}
\label{PT3}
\begin{split}
	\epsilon_1(R) 
	& = \frac{1}{8\pi} \sum_{abcd}^{\text{occ}}  \sum_{rs}^{\text{virt}} \frac{\langle ab \Vert rs \rangle \langle cd \Vert ab \rangle \langle rs \Vert cd \rangle }{\kappa_{a,b,r,s} \kappa_{c,d,r,s}}
	\\
	& + \frac{1}{8\pi} \sum_{ab}^{\text{occ}} \sum_{rstu}^{\text{virt}} \frac{\langle ab \Vert rs \rangle \langle rs \Vert tu \rangle \langle tu \Vert ab \rangle }{\kappa_{a,b,r,s}\kappa_{a,b,t,u}}
	\\
	& + \frac{1}{\pi} \sum_{abc}^{\text{occ}} \sum_{rst}^{\text{virt}} \frac{\langle ab \Vert rs  \rangle \langle cs \Vert tb  \rangle \langle rt \Vert ac \rangle}{\kappa_{a,b,r,s}\kappa_{a,c,r,t}}
	\\
	& + \frac{1}{\pi} \sum_{abc}^{\text{occ}}  \sum_{rst}^{\text{virt}} \frac{\langle ab \Vert rs  \rangle \langle ar \Vert ct  \rangle\langle rs \Vert ab  \rangle }{\kappa_{a,b,r,s}^2},
\end{split}
\end{equation}
with $\kappa_{a,b,r,s} =\kappa_a +\kappa_b - \kappa_r - \kappa_s$. Equation \eqref{PT3} can be decomposed, using the same transformations as in \eqref{eps-a-final} and  \eqref{eps-b-final}, into eight distinct contributions
\begin{equation}
	\epsilon_1(R) = \frac{1}{\pi^4} \sum_{k=1}^{8} \Xi_i(R).
\end{equation}
The explicit expressions of the $\Xi_k(R)$'s and their regions of integration are given in Table \ref{tab:epsilon1}.

\begin{table*}
	\caption{
	\label{tab:epsilon1}
	Explicit expressions of the $\Xi_k(R)$'s and their regions of integration. $v_R(q) = K_0(|q|R)$ and $\int d^mp = \int \ldots \int dp_1 \ldots dp_m$.}
	\begin{ruledtabular}
	\begin{tabular}{cll}	
	$\Xi_k(R)$			&	Integral	
							&	Region of integration
	\\
	\hline
	$\Xi_1(R)$			&	$\displaystyle \int dq\,d^3p \frac{v_R(q)^3}{q(p_1+p_2+q)q(p_1+p_3+q)}$
							&	$|p_i|<1$, $|p_i+q|>1$
	\\[12pt]
	$\Xi_2(R)$			&	$\displaystyle -\int dq\,d^3p\frac{v_R(q)^2 v_R(p_1-p_2)}{q(p_1+p_3+q)q(p_2+p_3+q)}$
							&	$|p_i|<1$, $|p_i+q|>1$
	\\[12pt]
	$\Xi_3(R)$			&	$\displaystyle -2\int dq\,d^3p \frac{v_R(q)^2v_R(p_1+p_2+q)}{q(p_1+p_2+q)q(p_1+p_3+q)}$
							&	$|p_i|<1$, $|p_i+q|>1$
	\\[12pt]
	$\Xi_4(R)$			&	$\displaystyle -2\int dq\,d^3p \frac{v_R(q)v_R(p_1+p_2+q)v_R(p_2-p_3)}{q(p_1+p_2+q)q(p_1+p_3+q)}$
							&	$|p_i|<1$, $|p_i+q|>1$
	\\[12pt]
	$\Xi_5(R)$			&	$\displaystyle -\int d^2q\,d^2p \frac{v_R(q_1)v_R(q_2)\left[v_R(q_1-q_2)-v_R(p_1-p_2)\right]}{q_1(p_1-p_2)q_2(p_1-p_2)}$
							&	$|p_1+q_1|>1$, $|p_1+q_2|>1$, $|p_1|<1$
	\\[12pt]
							&	
							&	$|p_2+q_1|<1$, $|p_2+q_2|<1$, $|p_2|>1$
	\\[12pt]
	$\Xi_6(R)$			& 	$\displaystyle \frac{1}{2}  \int d^2q\,d^2p \frac{v_R(q_1)v_R(q_2)\left[v_R(q_1-q_2)-v_R(p_1+p_2+q_1+q_2)\right]}{q_1(p_1+p_2+q_1)q_2(p_1+p_2+q_2)}$
							&	$|p_1+q_1|>1$, $|p_1+q_2|>1$, $|p_1|<1$
	\\[12pt]
							& 	
							&	$|p_2+q_1|>1$, $|p_2+q_2|>1$, $|p_2|<1$
	\\[12pt]
							& 	
							&	$|p_1+q_1|<1$, $|p_1+q_2|<1$, $|p_1|>1$
	\\[12pt]
							& 	
							&	$|p_2+q_1|<1$, $|p_2+q_2|<1$, $|p_2|>1$
	\\[12pt]
	$\Xi_7(R)$			&	$\displaystyle \int dq\,d^3p \frac{v_R(q)^2\left[v_R(p_1-p_2)-v_R(p_1-p_2+q)\right]}{(q^2+q(p_1+p_3))^2}$
							&	$|p_i|<1$, $|p_1+q|>1$, $|p_3+q|>1$
	\\[12pt]
	$\Xi_8(R)$			&	$\displaystyle \int dq\,d^3p \frac{v_R(q)v_R(p_1+p_2+q)\left[v_R(p_1-p_3+q)-v_R(p_1-p_3)\right]}{(q^2+q(p_1+p_2))^2}$
							&	$|p_i|<1$, $|p_1+q|>1$, $|p_2+q|>1$
	\\[12pt]
	\end{tabular}
	\end{ruledtabular}
\end{table*}

Again, for $R=0$, most of the integrals diverge. The first five terms have to be considered together, as well as the last two integrals while the sixth integral is finite.  Evaluating numerically each contribution and extrapolating the result to $R=0$ using the relation $\alpha R^{\beta} + \epsilon_1$ (see Fig.~\ref{fig:conv}), we find
\begin{equation}
\label{eps1}
	\epsilon_1 = \lim_{R \to 0} \epsilon_1(R) = +0.00844(7),
\end{equation}
which is agreement with the exact numerical value ($+0.008446$) obtained for the ring geometry of Ref.~\onlinecite{nQR12}. The error in \eqref{eps1} has been obtained by taking into account each numerical error estimate and extrapolating the overall error to $R=0$. \cite{Lee11b} We note that the present 1D UEG is one of the few systems where the $r_s$ coefficient of the high-density expansion is known. \cite{Endo99, Sun10} 
\begin{table}
	\caption{
	\label{tab:Ec}
	Reduced correlation energy ($-\Ec(r_s)$ in millihartree) for various $r_s$. The DMC results are computed using the CASINO software \cite{CASINO}  and are taken from Refs.~\onlinecite{Lee11a} and \onlinecite{nQR12}. Subscripts represent the statistical errors in the last digits. The deviation with respect to the DMC result is given in parenthesis.}
	\begin{ruledtabular}
	\begin{tabular}{clll}	
	$r_s$	&	DMC				&	This work				\\
	\hline
	0		&	---				&	27.416				\\
	0.2		&	$25.91_1$		&	25.90 ($-0.01$)		\\
	0.5		&	$23.962_1$		&	24.021 ($+0.059$)		\\
	1.0		&	$21.444\,1_2$		&	21.518 ($+0.074$)		\\
	2.0		& 	$17.922\,02_7$	&	17.927 ($+0.005$)		\\
	5.0		&	$12.317\,74_2$	&	12.220 ($-0.097$)		\\
	10.0		&	$8.292\,096_9$	&	8.201 ($-0.092$)		\\
	15.0		&	$6.319\,404_4$	&	6.251 ($-0.069$)		\\
	20.0		&	$5.132\,504_2$	&	5.081 ($-0.052$)		\\
	\end{tabular}
	\end{ruledtabular}
\end{table}

In summary, we have shown that the high-density correlation energy expansion \eqref{Ecjellium} of the 1D UEG is 
\begin{equation} 
\label{Ec_high}
	\Ec(r_s) = -\frac{\pi^2}{360} + 0.00845\,r_s + \ldots.
\end{equation}
We note that, contrary to the 2D and 3D UEGs, the expansion \eqref{Ec_high} does not contain any logarithm term up to first order in $r_s$, i.e. $\lambda_0 = \lambda_1 = 0$ (cf Eq.~\eqref{Ecjellium}).

\begin{figure}
\includegraphics[width=0.4\textwidth]{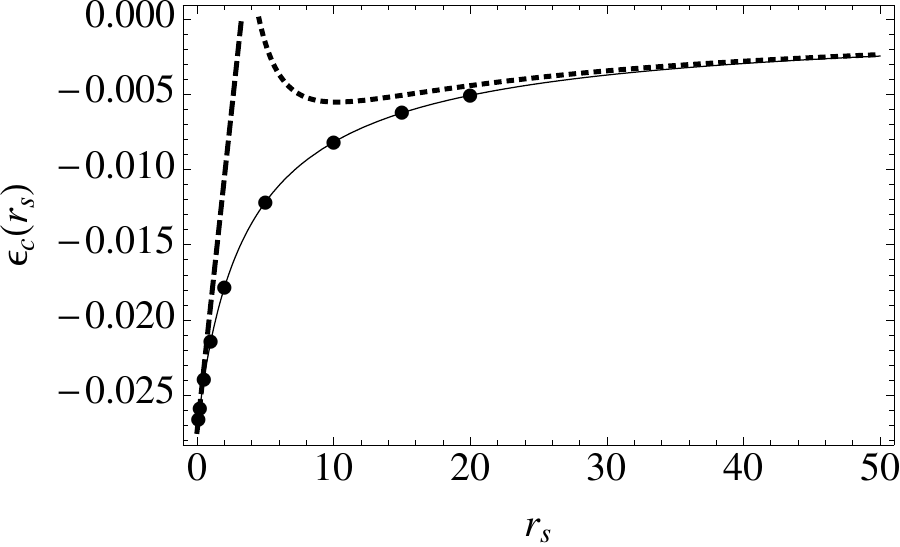}
\caption{
\label{fig:Ec}
$\Ec(r_s)$ given by Eq.~\eqref{Ec-Cios}  as a function of $r_s$ (solid line). DMC results are shown by black dots. The small-$r_s$ expansion of Eq.~\eqref{Ec_high} (dashed line) and large-$r_s$ approximation of Eq.~\eqref{Ec_low} (dotted line) are also shown.}
\end{figure}

\section{LDA functional}

For the 1D UEG, it is known \cite{Fogler05a, nQR12} that the low-density (large-$r_s$) expansion of the correlation energy is
\begin{equation}
\label{Ec_low}
\begin{split}
	\Ec(r_s) 
	& = \frac{\eta_0}{r_s} + \frac{\eta_1}{r_s^{3/2}} + \ldots
	\\
	& = -\frac{\ln(\sqrt{2\pi})-3/4}{r_s} + \frac{0.359933}{r_s^{3/2}} + \ldots.
\end{split}
\end{equation}
Using the ``robust'' interpolation proposed by Cioslowski \cite{Cioslowski12} and the high- and low-density expansions \eqref{Ec_high} and \eqref{Ec_low},  the correlation energy can be approximated by
\begin{equation}
\label{Ec-Cios}
	\Ec^{\text{LDA}}(r_s) = t^2 \sum_{j=0}^{3} c_j t^j (1-t)^{3-j},
\end{equation}
with 
\begin{equation}
	t = \frac{\sqrt{1+4\,k\,r_s}-1}{2\,k\,r_s},
\end{equation}
and
\begin{align}
	c_0 & = k\,\eta_0,	
	&
	c_1 & = 4\,k\,\eta_0+k^{3/2}\eta_1,
	\\
	c_2 & = 5\,\epsilon_0+\epsilon_1/k,
	&
	c_3 & = \epsilon_1,
\end{align}
where $k=0.414254$ is a scaling factor which is determined by a least-square fit of the DMC data given in Refs.~\onlinecite{Lee11a} and \onlinecite{nQR12}. 

We disagree with the last comment made in Ref.~\onlinecite{Cioslowski12}, which claims that this type of interpolation is not applicable to cases where the high- and low-density asymptotic expansions pertain to {\em de facto} different states, e.g. the 3D UEG. We claim that the non-applicability of such an interpolation is only due to the presence of logarithmic terms in the 2D and 3D UEGs. However, in our case, the 1D UEG does not involve any non-analytical terms. Thus, the methodology of Ref.~\onlinecite{Cioslowski12} is applicable in the present case.

The results using the new correlation functional \eqref{Ec-Cios} are compared to the DMC calculations of Refs.~\onlinecite{Lee11a} and \onlinecite{nQR12}. The results are gathered in Table \ref{tab:Ec} and depicted in Fig.~\ref{fig:Ec}.  For $0.2 \le r_s \le 20$, the LDA and DMC correlation energies agree to within 0.1 millihartree, which is remarkable given the simplicity of the functional. Overall, our LDA correlation functional gives accurate estimates of the correlation energy. 

\section{Conclusion}

In this Regular Article, we have shown that the expression of the high-density correlation energy for the 1D UEG is $\Ec(r_s) = -0.02742 + 0.00845 r_s+\ldots$. Combining these new results with the low-density correlation energy expansion $\Ec(r_s) = -[\ln(\sqrt{2\pi})-3/4]\,r_s^{-1} + 0.359933\,r_s^{-3/2}+\ldots$ and the available DMC data, we have proposed a LDA correlation functional, which yields satisfactory estimates of the correlation energy at high, intermediate and low densities. We believe these new results will be valuable for electronic structure calculations (especially within DFT).

\begin{acknowledgements}
The author thanks Peter Gill and Joshua Hollett for many fruitful discussions, the NCI National Facility for a generous grant of supercomputer time and the Australian Research Council for a Discovery Early Career Researcher Award (Grant DE130101441).
\end{acknowledgements}

\end{document}